# Microsecond non-melt UV laser annealing for future 3D-stacked CMOS

Toshiyuki Tabata,[a] Fabien Rozé, Louis Thuries, Sebastien Halty, Pierre-Edouard Raynal, Karim Huet and Fulvio Mazzamuto

*Laser Systems & Solutions of Europe (LASSE), 145 Rue Des Caboeufs, 92230 Gennevilliers, France*

Abhijeet Joshi and Bulent M. Basol

*Active Layer Parametrics (ALP), 5500 Butler Lane, Scotts Valley, 95066 California, USA*

Pablo Acosta Alba and Sébastien Kerdilès

*Université Grenoble Alpes, CEA-Leti, Grenoble, 38000, France*

a) Electronic mail: toshiyuki.tabata@screen-lasse.com

**ABSTRACT**

Three-dimensional (3D) CMOS technology encourages the use of UV laser annealing (UV-LA) because the shallow absorption of UV light into materials and the process timescale typically from nanoseconds (ns) to microseconds (µs) strongly limit the vertical heat diffusion. In this work, µs UV-LA solid phase epitaxial regrowth (SPER) demonstrated an active carrier concentration surpassing $1 \times 10^{21}$ at./cm$^3$ in an arsenic ion-implanted silicon-on-insulator substrate. After the subsequent ns UV-LA known for improving CMOS interconnect, only a slight (~5%) sheet resistance increase was observed. The results open a possibility to integrate UV-LA at different stages of 3D-stacked CMOS.

**Keywords:** laser anneal, dopant activation, process simulation, differential Hall effect metrology.

**MANUSCRIPT**

Three-dimensional (3D) integration of transistors[1-8] enables further increment of chip density and performance improvement while alleviating scaling difficulties. Applicable thermal budget is then severely restricted because the bottom-layer transistors must be preserved during the thermal processing of the upper-layer ones. Although the formation of junctions in two-dimensionally scaled transistors generally requires high temperature processing (e.g., ~1000 °C for a few seconds, heating up the wafer in its whole thickness), it becomes no longer acceptable for 3D-integrated transistors.[9]

UV laser annealing (UV-LA) can be a solution because the shallow absorption of UV light into semiconductor materials and the process timescale typically from nanoseconds (ns) to microseconds (μs) strongly limit the vertical heat diffusion.[1,6,10,11] Previous works shed light on the benefit of melt LA for metastable activation of dopants.[8,12-15] However, the reachable activation level seems dominated by the solidification velocity[14] so that locally varying heat dissipation in a real device structure may strongly affect it.[16] In addition, melting of doped Si degrades surface morphology.[3,14] These aspects might be a hindrance of integrating melt LA into a transistor fabrication flow. On the other hand, LA-induced solid phase epitaxial regrowth (SPER)[13,17-19] also demonstrated metastable activation of dopants.[20-21] In SPER, the temperature is one of the factors that determine the regrowth rate.[22] In addition, it can be enhanced by the presence of dopants at the moving amorphous/crystalline (a/c) interface.[23] One may therefore speculate that the time-temperature profile during LA SPER and the doping profile dominantly affect the finally obtained active carrier profile. In fact, millisecond LA SPER is well investigated in arsenic (As) ion-implanted Si, including the thermal stability of the activated As atoms.[18,20,24] However, it utilizes an argon-ion continuous-wave laser,[18] whose wavelength is typically in a near infra-red range,[25] and is not as efficient as UV-LA to limit the vertical heat diffusion.

We previously reported μs UV-LA SPER with an As-implanted silicon-on-insulator (SOI) substrate.[26] Although SPER was evidenced by physical analysis, there were some remaining issues such as the process non-uniformity represented by surface roughness and the inaccurate evaluation of active carrier concentration by electrochemical capacitance voltage profiling.[27] In this work, we used a modified μs UV-LA, which provides a top-hat beam profile instead of a Gaussian one, to improve the surface roughness after SPER. The active carrier concentration was then evaluated by the differential Hall effect methodology (DHEM), which uses a Van der Pauw pattern coupled with an electrochemical stripping process to measure the carrier concentration depth profile of a semiconductor material.[28-29] In addition, the thermal stability of the activated As atoms was investigated to assess the possibility of integrating UV-LA into different stages of 3D-integrated transistors. In fact, LA has already been successfully integrated into the back end of lines (BEOL), outperforming conventional furnace anneals.[30-32]

Silicon-on-insulator (SOI) wafers were used as the substrate. The top (100) silicon (Si) layer thickness was 70 nm, whereas that of the buried Si dioxide ($SiO_2$) layer was 145 nm. Arsenic ion implantation (I/I) was performed at room temperature (RT) at 19 keV with a dose of $4 \times 10^{15}$ at./cm$^2$. A 37-nm-thick amorphization from the Si surface was observed by cross-sectional transmission electron microscopy (TEM). The wafers were submitted to μs UV-LA at RT under a nitrogen ($N_2$) flow for As activation and crystal regrowth by SPER, enabling the different time-temperature profiles (Processes A and B) by means of the irradiated laser fluence and dwell time to assess their impact on the activation and redistribution of the As atoms.

To investigate the thermal stability of the activated As atoms, ns UV-LA followed at RT under a $N_2$ flow, accumulating single pulse irradiation as reported in Ref. 32 (e.g., 1, 10, 100, and 1000 times). The laser wavelength of both µs and ns UV-LA was between 300 and 400 nm. The sheet resistance ($R_{sq}$) was measured by a standard four-point probe. The As chemical concentration profile was measured by secondary ion mass spectroscopy (SIMS). The active As concentration profile was extracted by DHEM (ALPro™100, Active Layer Parametrics, Inc.). The surface roughness was evaluated by atomic force microcopy (AFM) with a 5 µm × 5 µm scan area, using the root-mean-square (RMS) value. A self-consistent time-harmonic solution based on the Maxwell equations allowed to simulate the time profile of the temperature ramping-up and cooling-down induced by UV-LA.[33] The dwell time was defined as the "full width at half maximum" of each profile.

Figure 1 shows the simulated µs UV-LA SPER time-temperature profiles. Both Processes A and B have a maximum temperature ($T_{max}$) close to 1000 °C without exceeding the melting point of a-Si (1420 K in Ref. 34). The absence of oxygen incorporation (i.e., non-melt) and the monocrystalline regrowth were re-confirmed as in Ref. 26 (data not shown). The dwell time is 8.3 µs and 12.7 µs for Processes A and B, respectively. The measured $R_{sq}$ values were equivalent (107.9 and 109.3 ohm/sq with an error of ± 1.5 %). Processes A and B gave a greater RMS value (0.18 and 0.20 nm) than the as-implanted surface (0.10 nm), but these are much reduced compared to the previous µs UV-LA SPER (1.5 nm).[26]

Figure 2 shows the As SIMS profiles taken before and after µs UV-LA SPER. Both Processes A and B introduced As migration towards the surface. The condensation of impurities at the a/c interface during SPER may explain it.[35] This As surface migration may be beneficial for lowering the contact resistivity of transistors.[13,36] The As chemical concentration at 20 nm depth is $1.25 \times 10^{21}$ at./cm$^3$ for Process A, whereas $0.98 \times 10^{21}$ at./cm$^3$ for Process B. Their difference is greater than the measurement error (± 10%). The slower ramping-up of Process B (~60 °C/µs) than Process A (~80 °C/µs) might have facilitated the dopants to migrate. By contrast, as expected from the process timescale and the diffusion coefficient of As in c-Si (~$10^{-14}$ cm$^2$/s at ~1000 °C from Ref. 37), the as-implanted As profiles were maintained in the non-amorphized SOI after Processes A and B. Figure 2 also shows the DHEM profiles taken after µs UV-LA SPER. The $R_{sq}$ value measured after Processes A and B by the DHEM equipment was 100.7 ohm/sq and 106.4 ohm/sq with an error of less than ± 1.0%, respectively. In both cases, the active level in the initial 30-nm-thick layer surpassed the As solid solubility in c-Si at ~1000 °C (~$3 \times 10^{20}$ at./cm$^3$ in Ref. 24) and became higher than $1 \times 10^{21}$ at./cm$^3$ near the surface. The activation ratio calculated from the SIMS and DHEM profiles (excluding the initial few nanometers because of the presence of a native oxide) was ~76% and ~73% for Processes A and B, respectively. The inactive As could be associated with the formation of deep levels ($As_nV$ (n = 2, 3, and 4) clusters, where V stands for vacancies in c-Si).[38] In the regrown SOI, V might be generated due to the rapid placement of atoms at the moving a/c interface. In the non-amorphized SOI, V should be related to the I/I damage not fully cured during µs UV-LA. Although $As_nV$ grows with time[39] and certainly rules the As deactivation, discussing its growth mechanism is beyond the scope of this paper.

The thermal stability of the activated As atoms has been assessed, assuming the typical BEOL anneal conditions (100 to 420 °C for 10 m to 1 h with furnace anneal,[40-43] whereas ~1300 °C for ~$10^{-7}$ to ~$10^{-4}$ s with LA).[30,43] In Ref. 24, the As deactivation in Si is discussed, starting from an active level (~$1 \times 10^{21}$ at./cm$^3$) similar to our µs UV-LA SPER cases. Then, an Arrhenius plot relevant to a 7% loss of sheet concentration is conceived to extract the activation energy of the As deactivation, which is found to be 2.0 eV and smaller than that of the SPER in intrinsic Si (2.7 eV in Ref. 22). Although the studied temperature range is limited between 350 and 410 °C, it would be an interesting attempt to extrapolate this Arrhenius plot towards a higher temperature range and plot the typical BEOL annealing conditions together. Figure 3 shows this benchmarking, indicating that most of the presented copper (Cu) BEOL anneals would maintain the high As active level, whereas most of the ruthenium (Ru) ones might lead to non-negligible active As loss.

The deactivation anneal (DA) was performed by ns UV-LA. Figure 4 shows the simulated ns UV-LA time-temperature profiles, where $T_{max}$ is ~1000 °C (Process C) or ~1300 °C (Process D) without exceeding the melting point of c-Si (1690 K in Ref. 44). Figure 5 shows the $R_{sq}$ evolution as a function of the accumulated DA time. SIMS confirmed no As redistribution during ns UV-LA DA (data not shown). The As deactivation is more pronounced with Process C than Process D. The As solid solubility in c-Si can explain it (~$3 \times 10^{20}$ at./cm$^3$ at ~1000 °C, whereas ~$6 \times 10^{20}$ at./cm$^3$ at ~1200 °C (possibly a bit higher at ~1300 °C)).[20] However, Processes C and D reach the same $R_{sq}$ degradation level as the DA time increases. This must be carefully examined, considering a complex system involving the As diffusion assisted by V or interstitials and the formation energy of various clusters such as $As_n$, $As_nV$, and $As_nV_m$.[38]

In summary, the µs UV-LA SPER was investigated to activate the As atoms implanted in the SOI layer. Prior to µs UV-LA, the 37-nm-thick amorphization was introduced in the 70-nm-thick SOI. Two µs UV-LA SPER conditions (Processes A and B) targeting $T_{max}$ close to 1000 °C with different dwell times were applied. Both showed the high active carrier concentration surpassing $1 \times 10^{21}$ at./cm$^3$ near the regrown SOI surface, accompanied by the As surface migration, which is beneficial for lowering the contact resistivity of transistors. The thermal stability of these activated As atoms was assessed by the ns UV-LA DA, where $T_{max}$ was set at ~1000 °C (Process C) or ~1300 °C (Process D), considering the typical BEOL LA conditions. The maximum $R_{sq}$ degradation ratio was ~5 % in the studied DA timescale, encouraging UV-LA integration into different stages of a 3D-stacked transistor fabrication flow to boost chip performance further. Although the level of active carrier concentration achieved in this work meets the current requirement of the state-of-the-art CMOS technologies,[45-46] the formation of donor-V complexes (e.g., $As_nV$) might restrict its additional enhancement. Then, the use of alternative doping elements, especially the chalcogens such as selenium[47] and tellurium,[48] may provide a solution. Although these elements are knowns as deep-level impurities, their increasing chemical concentration in Si triggers insulator-to-metal transition and allows non-saturating free-electron generation.

**ACKNOWLEDGMENTS**

The work covered by LASSE in this paper was supported by the IT2 project. This project has received funding from the ECSEL Joint Undertaking (JU) under grant agreement No 875999. The JU receives support from the European Union's Horizon 2020 research and innovation programme and Netherlands, Belgium, Germany, France, Austria, Hungary, United Kingdom, Romania, Israel.

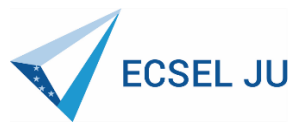
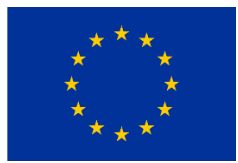

**FIGURES**

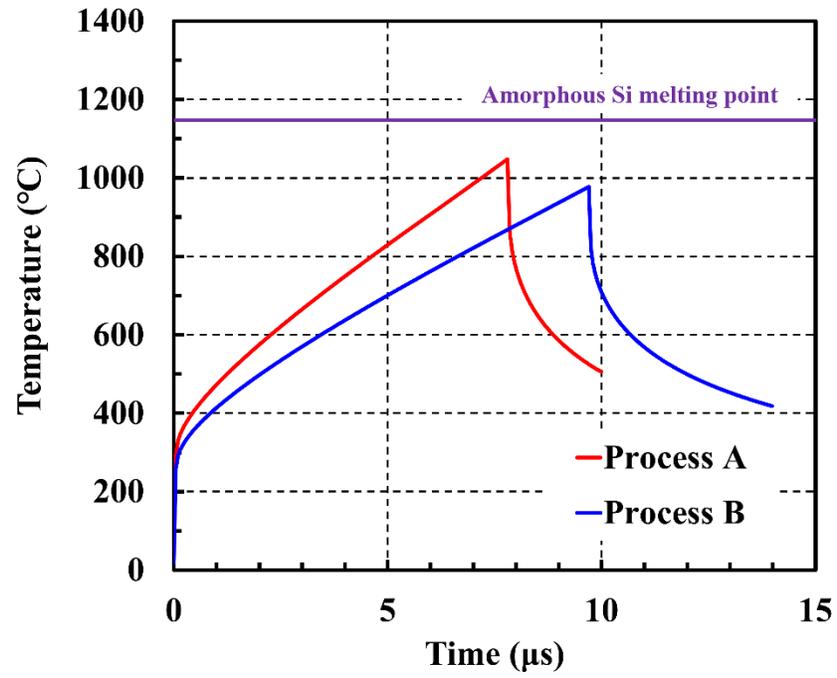

**FIG. 1.** UV-LA SPER time-temperature profiles obtained by simulation. The dwell time is 8.3 µs and 12.7 µs for Processes A and B, respectively. The a-Si melting point (1420 K in Ref. 34) is also shown.

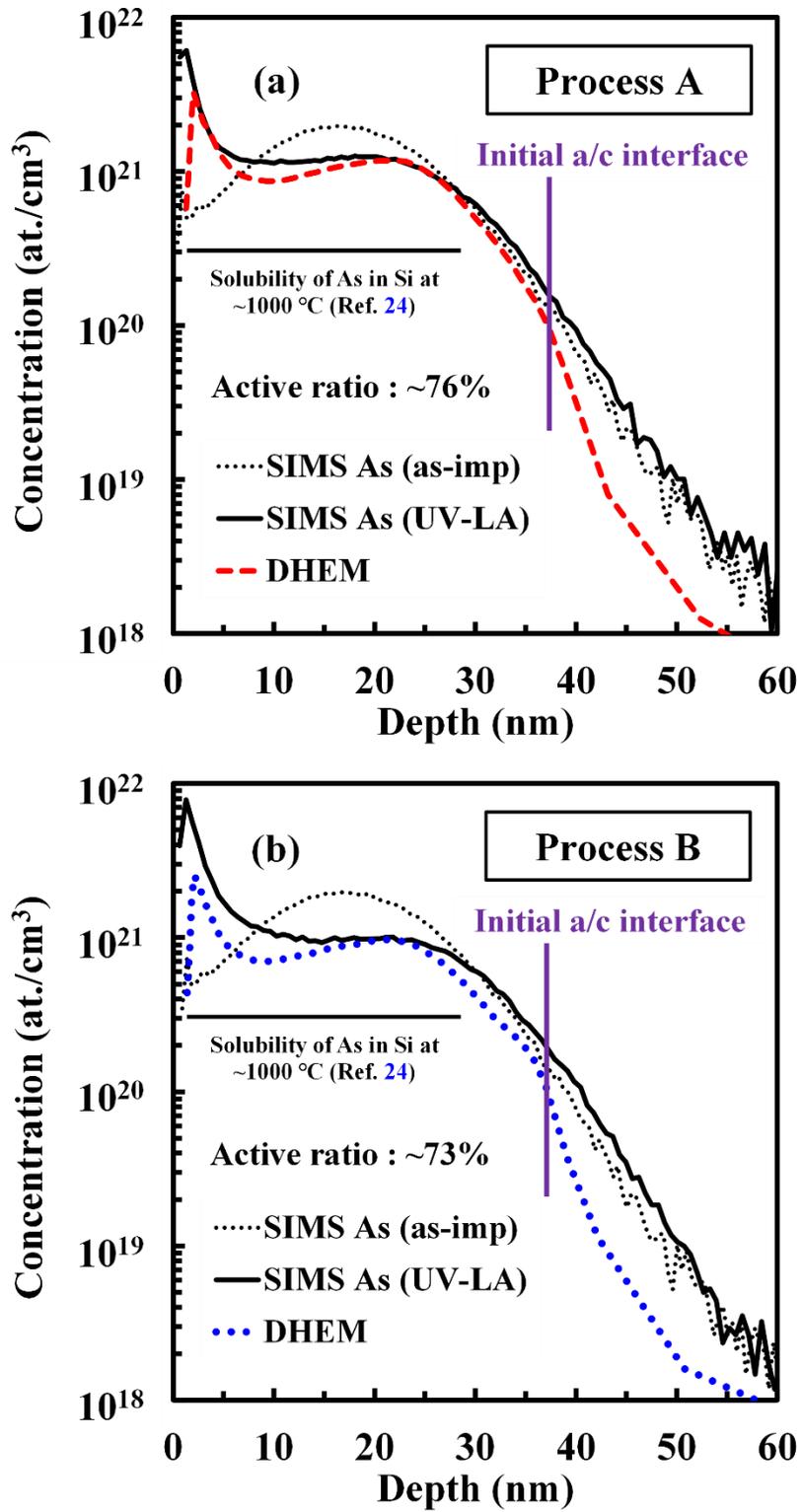

**FIG. 2.** SIMS and DHEM profiles taken after Processes A and B. The initial a/c interface position and the As solid solubility in c-Si at ~1000 °C reported in Ref. 24 are also indicated.

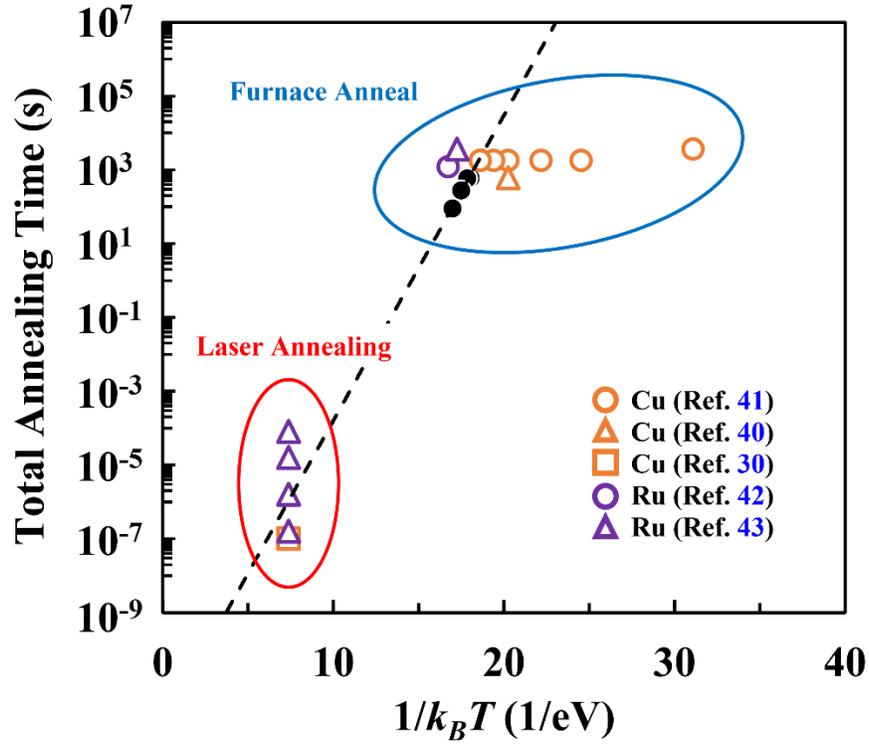

**FIG. 3.** Benchmarking of the typical BEOL anneal conditions against the Arrhenius plot relevant to a 7% loss of sheet concentration in As-doped Si after LA SPER giving an active level of ~1 × $10^{21}$ at./cm³ (see Ref. 24).

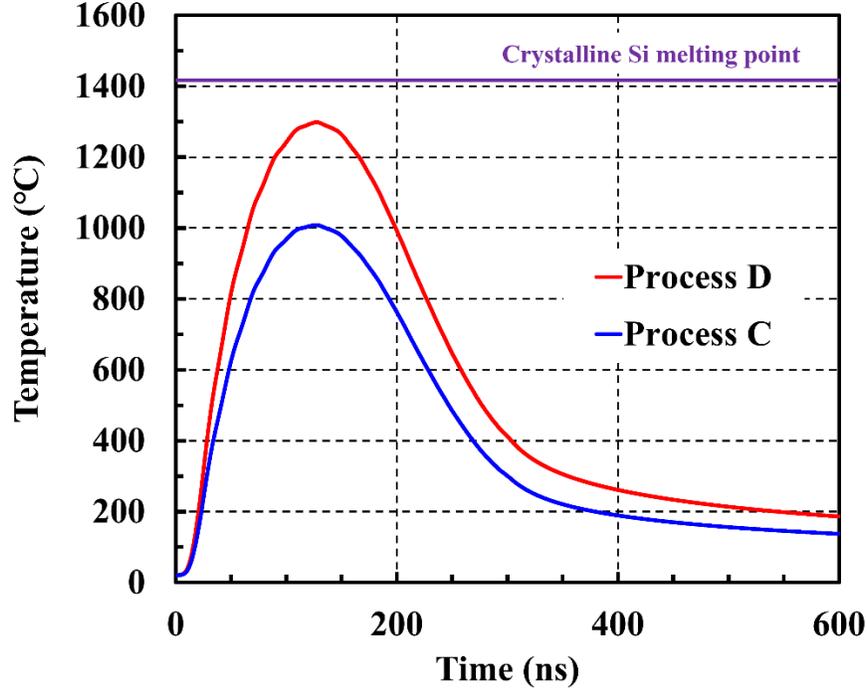

**FIG. 4.** UV-LA DA time-temperature profiles obtained by simulation. The dwell time is ~2 × $10^2$ ns for both Processes C and D. The c-Si melting point (1690 K in Ref. 44) is also shown.

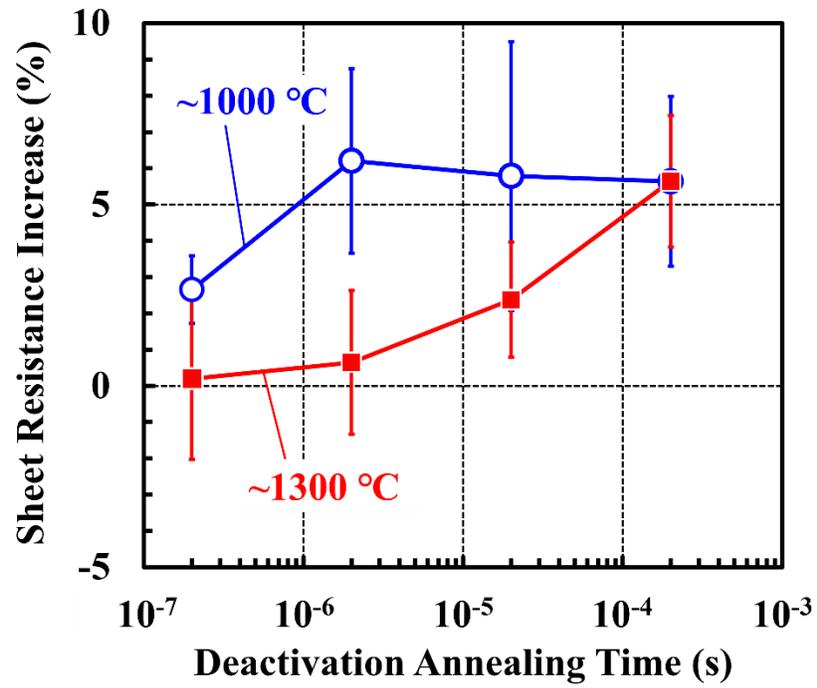

**FIG. 5.** $R_{sq}$ degradation ratio as a function of the accumulated DA time. The error bars come from the sample uniformity and $R_{sq}$ measurement accuracy.